# Prediction of Prokaryotic and Eukaryotic Promoters Using Convolutional Deep Learning Neural Networks

Victor Solovyev[1*], Ramzan Umarov[2]

[1] Softberry Inc., Mount Kisco, USA; [2] King Abdullah University of Science and Technology, Thuwal, KSA

[*] Corresponding author: victor@softberry.com

## Abstract

Accurate computational identification of promoters remains a challenge as these key DNA regulatory regions have variable structures composed of functional motifs that provide gene specific initiation of transcription. In this paper we utilize Convolutional Neural Networks (CNN) to analyze sequence characteristics of prokaryotic and eukaryotic promoters and build their predictive models. We trained the same CNN architecture on promoters of four very distant organisms: human, plant (Arabidopsis), and two bacteria (Escherichia coli and Mycoplasma pneumonia). We found that CNN trained on sigma70 subclass of Escherichia coli promoter gives an excellent classification of promoters and non-promoter sequences (Sn=0.90, Sp=0.96, CC=0.84). The Bacillus subtilis promoters identification CNN model achieves Sn=0.91, Sp=0.95, and CC=0.86. For human and Arabidopsis promoters we employ CNNs for identification of two well-known promoter classes (TATA and non-TATA promoters). CNNs models nicely recognize these complex functional regions. For human Sn/Sp/CC accuracy of prediction reached 0.95/0.98/0,90 on TATA and 0.90/0.98/0.89 for non-TATA promoter sequences, respectively. For Arabidopsis we observed Sn/Sp/CC 0.95/0.97/0.91 (TATA) and 0.94/0.94/0.86 (non-TATA) promoters. Thus, the developed CNN models (implemented in CNNProm program) demonstrated the ability of deep learning with grasping complex promoter sequence characteristics and achieve significantly higher accuracy compared to the previously developed promoter prediction programs. As the suggested approach does not require knowledge of any specific promoter features, it can be easily extended to identify promoters and other complex functional regions in sequences of many other and especially newly sequenced genomes. The CNNProm program is available to run at web server http://www.softberry.com.

**Key words**: Convolutional neural networks, deep learning, DNA sequence analysis, Promoter prediction, animal, plant and bacterial promoters

**Introduction.**

Promoter is a key region that is involved in differential transcription regulation of protein-coding and RNA genes. The gene-specific architecture of promoter sequences makes it extremely difficult to devise the general strategy for their computational identification [1, 2]. Promoter 5'-flanking regions may contain many short (5–10 bases long) motifs that serve as recognition sites for proteins providing initiation of transcription as well as specific regulation of gene expression.

The minimal eukaryotic promoter region called the core promoter is capable of initiating basal transcription and contains a transcription start site (TSS). About 30–50% of all known eukaryotic promoters contain a TATA-box at a position ~30 bp upstream from the transcription start site. Many highly expressed genes contain a strong TATA box in their core promoter. At the same time, large groups of genes including housekeeping genes, some oncogenes and growth factor genes possess TATA-less promoters. In these promoters Inr (the initiator region) or the recently found downstream promoter element (DPE), usually located ~25-30 bp downstream of TSS, may control the exact position of the transcription start [1, 2].

Bacterial promoters contain two short conserved sequence elements approximately -10 and -35 nucleotides upstream from the transcription start site. The -10 box is absolutely essential to start transcription in prokaryotes. The sequence of -35 box affects the transcription rate [3-6]. Those consensus sequences, while conserved on average, are not found intact in most promoters.

Accurate prediction of promoters is fundamental for interpreting gene expression patterns, and constructing and understanding genetic regulatory networks. For the last decade genomes of many organisms have been sequenced and their gene content was mainly computationally identified, however, the promoters and transcriptional start sites (TSS) are still undetermined in most cases and the efficient software able to accurately predict promoters in newly sequenced genomes is not yet available in public domain.

There are many attempts to develop promoter prediction software as for bacterial as well as for eukaryotic organisms. Most of them implemented very different computational algorithms, which often account some specific sequence features discovered during experimental studies. Fickett and Hatzigeorgiou [7] presented one of the first reviews of eukaryotic promoter prediction programs. Among these were oligonucleotide content-based neural network and the linear discriminant approaches.

**Performance measures**

Several measures to estimate the accuracy of a recognition function were introduced in genomic research [8, 9]. Consider that we have S sites (positive examples) and N non-sites (negative examples). By applying the recognition function we identify correctly Tp sites (true positives) and Tn non-sites (true negatives). At the same time Fp (false positives) sites were wrongly classified as non-sites and Fn (false negative) non-sites were wrongly classified as sites. Sensitivity (Sn) (true positive rate) measures the fraction of the true positive examples that are correctly predicted: Sn=Tp/(Tp+Fn). Specificity (Sp) (true negative rate) measures the fraction of the predicted sites that are correct amongst those predicted: Sp=Tn/(Tn+Fp). Accuracy AC=(Tp+Tn)/(Tn+Tp+Fn+Fp) measures an average performance on positive and negative datasets. However, this measures does not take into account the possible difference in sizes of site and non-sites sets. A more correct single measure (correlation coefficient) takes the relation between correctly predictive positives and negatives as well as false positives and negatives into account [9]:

$$CC = \frac{(Tp \times Tn - Fp \times Fn)}{\sqrt{(Tp + Fp)(Tn + Fn)(Tp + Fn)(Tn + Fp)}}$$

It was shown that many general-purpose promoter prediction programs can typically recognize only ~50% of the promoters with a false positive (FP) rate of ~1 per 700–1000 bp [7]. The study to make a critical assessment of the human promoter prediction field also demonstrated a pretty low level of sensitivity of 58% for the specificity of 92% and correlation coefficient (CC) ranged from 0.52-0.73 for evaluated promoter predictors [10]. Much better accuracy has been observed for identification of plant promoters [11-16], however their specificity level does not exceed 90% that will generate significant number of false positives when applying to analyze long genomic sequences. The top two performers TSSP_TCM [11] and Promobot [12] with Sn=0.88-0.89 and Sp=0.84-0.86 outperform NNPP [13] (Sn/Sp:0.74/0.70), PromoterScan [14] (Sn/Sp:0.08/0.04), Promoter [15] (Sn/Sp:0.24/0.34), Prom-Machine [16] (Sn/Sp:0.86/0.81).

While bacterial promoters have simpler structure than transcription initiation regions of higher organisms, their identification is also represent a challenging task. Using sequence alignment kernel and SVM classifier Gordon et al. [17] achieved Sn=0.82, Sp=0.84 in recognition σ70 promoter and non-promoter E.coli sequences. Similar accuracy is observed for popular bacterial promoter prediction program Bprom [18]. These programs clearly outperform the NNPP (trained on E.coli K12 sequences) [13] and SIDD [19] programs. For example, SIDD correctly predicted 74.6% of the real promoters with a false positive rate of 18%. When NNPP correctly predicted 66.4% of the real promoters, its false positive rate is 22.4%.

Thousands genomes of bacteria and eukaryotic organisms are sequenced already and more will be sequenced soon, while little transcriptional information is available for most of them. Moreover, new genomes can have different promoter features than features observed during studying in model organisms. For example, recent studies have shown that TATA boxes and Initiators are not universal features of plant promoters, and that other motifs such as Y patches may play a major role in the transcription initiation in plants [12, 20-21]. We face the situation that specific promoter characteristics that often used in development of promoter predictors in many new genomes are poorly understood. This creates favorable circumstances for developing universally applicable algorithm of promoter prediction and in this paper we propose to use convolutional neural network with just sequence input as a rather general approach to solution of this problem.

Deep convolutional neural network is capable of achieving record-breaking results in processing images, video, speech and audio on highly challenging datasets using purely supervised learning and recently have won a large number of contests in pattern recognition and machine learning [22-25]. There are a few successful examples of applying them for biological problems. A deep learning–based algorithmic framework, DeepSEA, can predict chromatin effects of sequence alterations prioritize functional SNPs directly learning a regulatory sequence code from large-scale chromatin-profiling data, enabling prediction of chromatin effects of sequence alterations with single-nucleotide sensitivity [26]. Improved performance for this task reported using DanQ [27], a hybrid framework that combines convolutional and bi-directional long short-term memory recurrent [28-29] neural networks. Chen et al. applied deep learning method (abbreviated as D-GEX) to infer the expression of target genes from the expression of landmark genes [30].

In this paper we utilize Convolutional Neural Networks (CNN) to analyze sequence characteristics of prokaryotic and eukaryotic promoters and build their predictive models. The developed CNN models (implemented in CNNProm program) demonstrated the ability of deep learning to grasp complex promoter sequence characteristics and achieve significantly higher accuracy compared to the previously developed promoter prediction programs.

**Training and testing data**

In this study to demonstrate universality of the suggested approach to promoter prediction problem we selected promoter sequences from very distant group of organisms: two bacteria, human and one plant. The number of promoter and non-promoter sequences for each organism is presented in Table 1.

We used bacterial promoter and non-promoter sequences of size 81 nt (nucleotides). Bacterial non-promoter sequences were extracted from the corresponding genome sequences. From genome fragments that randomly located within coding regions we took their complementary chain sequences. Escherichia coli σ70 promoter sequences were extracted from the manually curated RegulonDB [36]. Bacillus subtilis promoters were taken from a collection described in [37]. As Human and Arabidopsis non-promoter sequences (size 251 nt) we used random fragments of their genes located after the first exon. Eukaryotic promoter sequences were extracted from the well-known EPD database [38].

Table 1. Number, length and location of promoter and non-promoter sequences for studied organisms. Location is relative to the TSS (Transcription Start Site) position.

| Organism | #promoter sequences | #non-promoter sequences | Length/Location |
|---|---|---|---|
| Escherichia coli σ70 | 839 | 3000 | 81/-60 - +20 |
| Bacillus subtilis | 746 | 2000 | 81/-60 - +20 |
| Human TATA | 1426 | 8256 | 251/-200 - +50 |
| Human non-TATA | 19811 | 27731 | 251/-200 - +50 |
| Arabidopsis TATA | 1497 | 2879 | 251/-200 - +50 |
| Arabidopsis non-TATA | 5905 | 11459 | 251/-200 - +50 |

We used 20% of each set sequences in our test sets. 90% of the remaining sequences were used as training and 10% as validation sets. Training sets provide data to generate parameters of CNNmodels and validation sets define the optimal number of learning epochs (cycles) that should be limited at some points to avoid over fitting.

## Results and Discussion

### CNN architecture for building promoter recognition models

There are many network architectures and the task is to construct a suitable one for a particular research problem. In *learnCNN.py* program we implemented CNN model using Keras - a minimalist, highly modular neural networks library, written in Python [32]. It uses Theano library [33-34] as backend and utilizes GPU [35] for fast neural network training. Adam optimizer is used for training with categorical cross-entropy as a loss function. Our CNN architecture (Fig. 1) in most cases consisted of just one convolutional layer with 200 filters having length 21. After convolutional layer, we have a standard Max-Pooling layer. The output from the Max-Pooling layer is fed into a standard fully connected ReLU layer. Pooling size was usually 2. Finally, the ReLU layer is connected to output layer with sigmoid activation, where neurons correspond to promoter and non-promoter classes. The batch size used for training was 16.

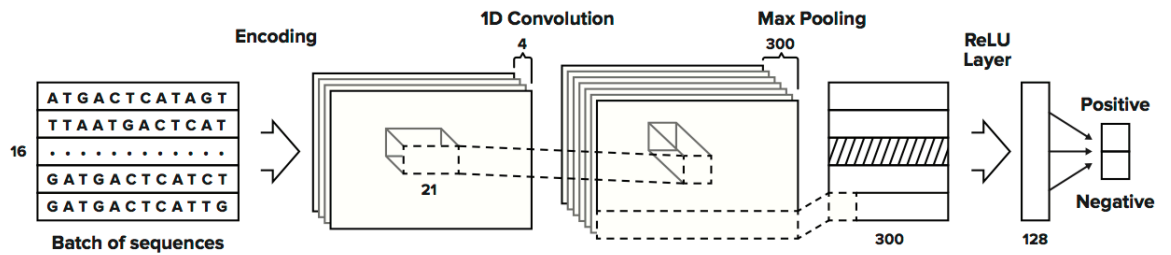

Fig.1. Basic CNN architecture that was used in building promoter models implemented in the *learnCNN.py* program (see description in the text).

Input of the network consist of nucleotide sequences where each nucleotide is encoded by 4 dimensional vector A (1,0,0,0), T(0,1,0,0), G(0,0,1,0) and C(0,0,0,1). Output is 2 dimensional vector: promoter (1, 0) and Non-promoter (0, 1) prediction. The training takes a few minutes with using GTX 980 Ti GPU. We intentionally used in most cases one layer CNN architecture, however to get a proper balance of accuracy between positives example (promoters) and negative examples (non-promoter) 2 or 3 layers was applied. A typical example of the model computation is presented in Fig. 2.

```
learncnn.py params_hu.txt
Using Theano backend.
Using gpu device 0: GeForce GT 650M (CNMeM is enabled with initial size: 16.0% of memory, cuDNN 5005)
Train on 6972 samples, validate on 775 samples
Epoch 1/5
6972/6972 [==========] - 6s - loss: 0.1705 - acc: 0.9389 - val_loss: 0.0570 - val_acc: 0.9768
Epoch 2/5
6972/6972 [==========] - 6s - loss: 0.0569 - acc: 0.9796 - val_loss: 0.0555 - val_acc: 0.9768
Epoch 3/5
6972/6972 [==========] - 6s - loss: 0.0207 - acc: 0.9933 - val_loss: 0.0496 - val_acc: 0.9755
Epoch 4/5
6972/6972 [==========] - 6s - loss: 0.0088 - acc: 0.9973 - val_loss: 0.0622 - val_acc: 0.9781
Epoch 5/5
6972/6972 [==========] - 6s - loss: 0.0031 - acc: 0.9991 - val_loss: 0.0794 - val_acc: 0.9755
1936/1937 [===============] - ETA: 0s ('Test score:', 0.10171617846821795)
('Test accuracy:', 0.9752194114610222)
('Sensitivity:', 0.9463087248322147)
('Specificity:', 0.9768151311775473)  ('CC:' 0.8968660492481778)
```

Fig.2. An example of learning CNN models for human promoters.

**The accuracy of promoter identification by constructed CNN models**

Using the CNN architecture described above implemented in ***learnCNN.py*** program we analyze the promoter and non-promoter sequence data (Table 1). The ***learnCNN.py*** learns parameters of CNN model and output the accuracy of Promoter Prediction for the test set of sequences. It also writes the computed CNN Model (PPCNNmodel) to the file that can be used later in programs for promoter identification in a given sequence.

The accuracy information and some parameters of CNN architecture used for the particular datasets are presented in Table 2.

Table 2.  The accuracy and parameters of CNN models.

| Organism | Sn | Sp | CC | CNN architecture[*] |
|---|---|---|---|---|
| Escherichia coli σ70 | 0.90 | 0.96 | 0.84 | 100, 7, 0 / 150, 21, 12 |
| Bacillus subtilis | 0.91 | 0.95 | 0.86 | 100, 15, 2 / 250, 17, 2 |
| Human TATA | 0.95 | 0.98 | 0.90 | 200, 21, 4 |
| Human non-TATA | 0.90 | 0.98 | 0.89 | 300, 21, 231 |
| Arabidopsis TATA | 0.95 | 0.97 | 0.91 | 200, 21, 4 |
| Arabidopsis non TATA | 0.94 | 0.94 | 0.86 | 200, 21, 2 |

[*] architecture as 200, 21, 4 describes one layer with 200 filters, 7 is the filter length and 4 is the pooling size; "/" separates two layers data.

We found that the computed CNN models demonstrated the ability of deep learning to grasp complex promoter sequence characteristics and achieve significantly higher accuracy compared to the previously developed promoter prediction programs. For example, CNN trained on sigma70 sub-class of Escherichia coli promoter give an excellent classification of promoters and non-promoter sequences (Sn=0.90, Sp=0.96).  For human and Arabidopsis promoters we employed CNNs for identification of two well-known promoter classes (TATA-box and non-TATA promoters). CNNs models nicely recognize these complex functional regions. For human Sn/Sp accuracy of prediction reached 0.95/0.98 for TATA-box and 0.90/0.98 for non-TATA promoter sequences, respectively. We observe outstanding performance for identification of Arabidopsis promoters as well:  Sp/Sn for TATA promoters 0.95/0.97 and for non-TATA promoters 0.94/0.94. It is very significant gain in prediction performance compared with evaluated human promoter predictors where the sensitivity of 58%, specificity of 92% and correlation coefficient (CC) ranged from 0.52-0.73 were observed [10].

We would like to point out an important benefit of the considered CNN models. While using as input just nucleotide sequences they can outperform recognition functions built based on preselected significant features. For example, widely used Bprom [18] promoter prediction program utilizes a set of seven features  (five relatively conserved sequence motifs, represented by their weight matrices, the distance between –10 and –35 elements and the ratio of densities of octa-nucleotides overrepresented in known bacterial transcription factor binding sites relative to their occurrence in the coding regions.  Computing these features for a set of 839 experimentally verified σ70 promoters from Regulon database [36] and 3000 non-promoter E.coli sequences and using LinearDiscriminantAnalysis and other discrimination approaches from scikit-learn Python library [39] we reached an average accuracy of classification of promoter and non-promoter sequences 0.92 applying cross-validation evaluation. The CNN model demonstrated a better recognition rate (Table 2) for the same data.

To apply our Promoter Prediction CNN (PPCNN) models to classify sequences into promoters and non-promoters we designed CNNprom.py program.  It takes the fasta format files as input together with the model parameters file and outputs classification results for each sequence. If the sequence is classified as promoter, the score assigned by network is provided in the output as well. The CNNProm program is available to run for sequences of four studied organisms at the Softberry web server http://www.softberry.com.

**Random substitution method to discover positionally conserved functional elements**

Analyzing network behavior we can extract some information on significant elements of the input data. Promoter sequences usually contain binding sites of regulatory proteins. Some of them occupy various locations relatively to TSS and can be found in direct or complementary DNA chain. However, there are a number of well-known functional sites (such as bacterial -10 –box or eukaryotic TATA-box) that occupy approximately the same position in each promoter sequence. To discover such sites we suggest the following procedure. Take a window of length L (including positions from x1 to x2) and change the sequence within this window to random sequence. Evaluate the accuracy of the site prediction after such change. Using sliding window going from the beginning of the functional site sequence we can build a performance profile that will reflect the effect of random sequence inserted in each sequence position instead of original sequence on the accuracy of the site prediction. An example of such profile computed with window size 6 nt is presented in Fig. 2.

We can see that substitution of the sequence located between -45 – -20 positions relatively to TSS of human promoters will drastically decrease the prediction accuracy. These positions include the well-known functional motif called TATA-box. The sequence logo showing consensus of motif is presented in Fig. 3.

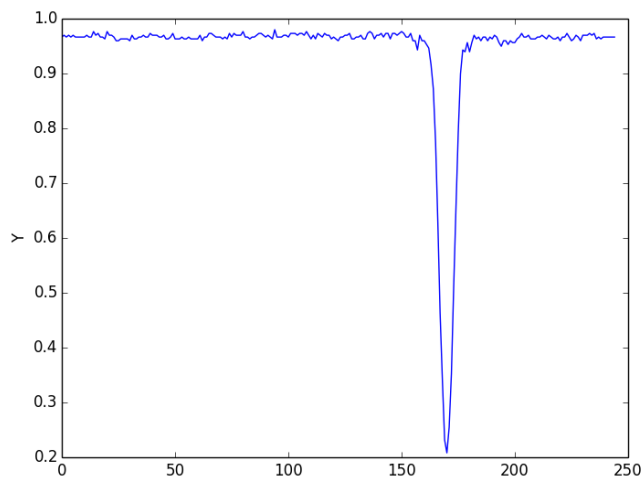

Fig. 2. Effect of random substitution of 6 nt sequence window on accuracy of TATA human promoters classification. Y axis is accuracy and X axis is window position.

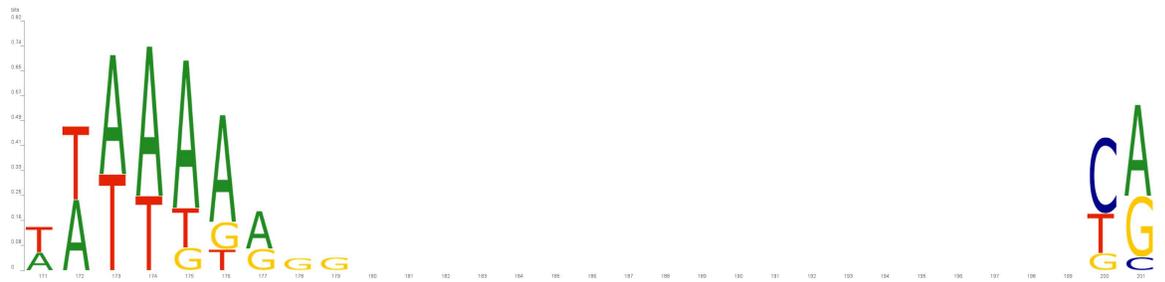

Fig.3. Sequence logo of human TATA promoter sequences in the TATA-box region and TSS region.

Another interesting example was observed in application of random substitution procedure to Arabidopsis non-TATA promoters (Fig.4).

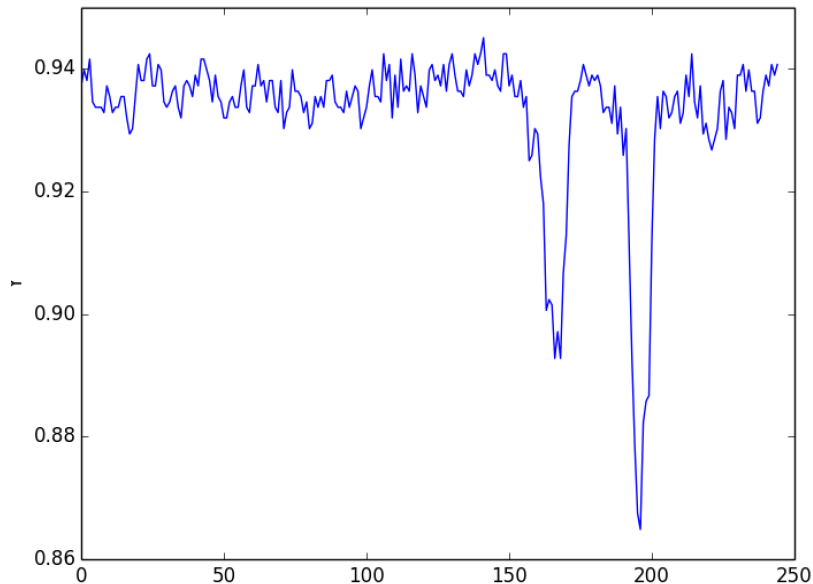

Fig. 4. Effect of random substitution of 6 nt sequence window on accuracy of non-TATA Arabidopsis promoters classification.

Here we observe two positionally conserved and potentially functionally important elements. One is located approximately in positions -34 - -28 and another in positions -2 – 0 relatively to TSS (position 0).

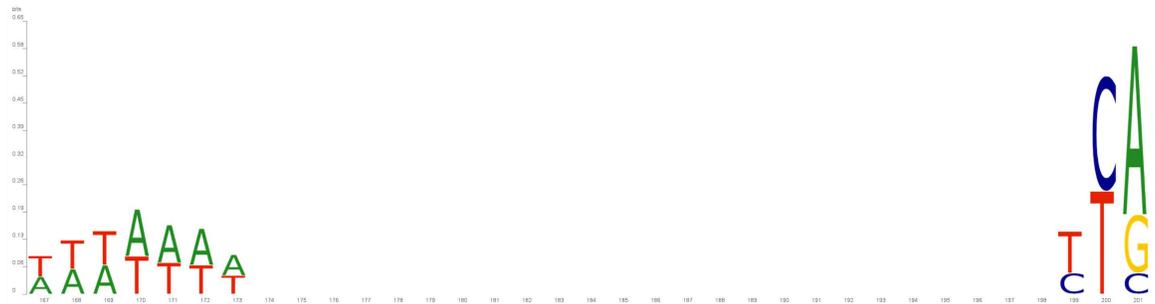

Fig.5. Sequence logo of Arabidopsis TATA promoter sequences in the TATA-box region and TSS region.

Thus, the suggested random substitution procedure can provide possibility to discover location of functionally important sites (sub-regions) that are still often poorly understood. Due to relatively high accuracy of CNNprom promoter prediction it would be interesting to use them in known or predicted upstream gene regions in combination with gene-recognition software tools to improve gene identification accuracy as well as make annotation of promoter regions.

We intentionally used a simple CNN architecture in this study, while it is possible to build more complicated models that can reach even better accuracy, but this paper aims to demonstrate the power of the approach.

The suggested application of deep learning in promoter prediction and positional functional sites analysis approach does not require knowledge of any specific promoter features. As the convolution filters able to automatically capture sequence motifs and other significant characteristics of biological/genomic sequences, this approach can be easily extended to identify promoters and other complex functional regions in sequences of many other and especially considering that complete genomic sequence of thousand organisms will soon be available and that little transcriptional information is known for most of them.

**References**

1. Sandelin A et al. (2007)_Mammalian RNA polymerase II core promoters: insights from genome-wide studies. *Nat Rev Genet.* 8(6):424-436.
2. Solovyev V, Shahmuradov I, Salamov A. (2010) Identification of promoter regions and regulatory sites. *Methods Mol Biol.* 674, 57-83.
3. Harley CB, Reynolds RP (1987) Analysis of E.coli promoter sequences. *Nucleic Acids Res.* 15: 2343-2361.
4. Lisser S, Margalit H (1993) Compilation of e.coli mrna promoter sequences. *Nucleic Acids Res.* 21: 1507-1516.
5. Jacques P., Rodrigue S., Gaudreau L., Goulet J., Brzezinski R. (2006) Detection of prokaryotic promoters from the genomic distribution of hexanucleotide pairs. BMC Bioinformatics 7,423.

6. Meysman P., Collado-Vides J., Morett E., Viola R., Engelen K., et al. (2014) Structural properties of prokaryotic promoter regions correlate with functional features. PLoS ONE 9: e88717.
7. Fickett J., Hatzigeorgiou A. (1997) Eukaryotic Promoter Recognition. Genome Res. 7, 861-878.
8. Burset M., Guigo R. (1996) Evaluation of gene structure prediction programs, Genomics, 34(3), 353-367.
9. Matthews B.W. 1975. Comparison of the predicted and observed secondary structure of T4 phage lysozyme. Biochem.Biophys.Acta 405: 442-451.
10. Bajic V., Brent M., Brown R., Frankish A., Harrow J., Ohler U., Solovyev V., Tan S. (2006) Performance assessment of promoter predictions on ENCODE regions in the EGASP experiment. *Genome Biol.* 7, Suppl 1, p. 3.1-3.13.
11. Shahmuradov I, Solovyev V. and Gammerman A. (2005) Plant promoter prediction with confidence estimation. *Nucleic Acids Research* 33(3),1069-1076.
12. Azad A., Shahid S., Noman N., Lee H. (2011) Prediction of plant promoters based on hexamers and random triplet pair analysis. Algorithms Mol Biol. 2011; 6: 19.
13. Reese M. (2001) Application of a time-delay neural network to promoter annotation in the Drosophila melanogaster genome. *Comput Chem* 26(1), 51-56.
14. Prestridge D. (1995) Predicting Pol II promoter sequences using transcription factor binding sites. J Mol Biol 249(5), 923-932.
15. Knudsen S. (1999) Promoter2.0: for the recognition of PolII promoter sequences. *Bioinformatics* 15(5), 356-361.
16. Anwar F., Baker M., Jabid T., Hasan M., Shoyaib M., Khan H., Walshe R. (2008) Pol II promoter prediction using characteristic 4-mer motifs: a machine learning approach. *BMC Bioinformatics* 9, 414.
17. Gordon L, Chervonenkis A, Gammerman A, Shahmuradov I, Solovyev V. (2003) Sequence alignment kernel for recognition of promoter regions. *Bioinformatics* 19(15), 1964-1971.
18. Solovyev V, Salamov A. (2011) Automatic Annotation of Microbial Genomes and Metagenomic Sequences. In *Metagenomics and its Applications in Agriculture, Biomedicine and Environmental Studies* (Ed. R.W. Li), Nova Science Publishers, p. 61-78.
19. Wang H. and Benham C. (2006) Promoter prediction and annotation of microbial genomes based on DNA sequence and structural responses to superhelical stress. BMC Bioinformatics 7, 248.
20. Yamamoto Y., Ichida H., Abe T., Suzuki Y., Sugano S., Obokata J. (2007) Differentiation of core promoter architecture between plants and mammals revealed by LDSS analysis. Nucleic Acids Res 35(18), 6219-6226.
21. Civan P., Svec M. (2009) Genome-wide analysis of rice (Oryza sativa L. subsp. japonica) TATA box and Y Patch promoter elements. Genome 52(3), 294-297.
22. Krizhevsky, A., Sutskever, I., & Hinton, G. E. (2012). Imagenet classification with deep convolutional neural networks. In *Advances in neural information processing systems* (pp. 1097-1105).

23. Christian Szegedy, Wei Liu, Yangqing Jia, Pierre Sermanet, Scott Reed, Dragomir Anguelov, Dumitru Erhan, Vincent Vanhoucke, Andrew Rabinovich (2015) Going Deeper With Convolutions The IEEE Conference on Computer Vision and Pattern Recognition (CVPR), 2015, pp. 1-9.
24. Yann LeCun, Yoshua Bengio & Geoffrey Hinton (2015) Deep learning. Nature 521, 436–444.
25. Schmidhuber, Jürgen. "Deep learning in neural networks: An overview." *Neural Networks* 61 (2015): 85-117.
26. Zhou J. ,Troyanskaya O. (2015) Predicting effects of noncoding variants with deep learning–based sequence model. Nat Methods. 12(10), 931–934.
27. Quang D, Xie X.(2016) DanQ: a hybrid convolutional and recurrent deep neural network for quantifying the function of DNA sequences. Nucleic Acids Res. 44(11), e107.
28. Graves,A. and Schmidhuber,J. (2005) Framewise phoneme classification with bidirectional LSTM and other neural network architectures. Neural Net., 18, 602–610.
29. Graves, A., Mohamed, A. R., & Hinton, G. (2013, May). Speech recognition with deep recurrent neural networks. In 2013 IEEE international conference on acoustics, speech and signal processing, IEEE, 6645-6649.
30. Chen,Y., Li,Y., Narayan,R., Subramanian,A. and Xie,X. (2016) Gene expression inference with deep learning. Bioinformatics, 2(12), 1832-1839.
31. Alipanahi,B., Delong,A., Weirauch,M.T. and Frey,B.J. (2015) Predicting the sequence specificities of DNA- and RNA-binding proteins by deep learning. Nat. Biotechnol., 33, 831–838.
32. Chollet F. (2015) Keras: Deep Learning library for Theano and TensorFlow. GitHub., https://github.com/fchollet/keras.
33. F. Bastien, P. Lamblin, R. Pascanu, J. Bergstra, I. Goodfellow, A. Bergeron, N. Bouchard, D. Warde-Farley and Y. Bengio. (2012) Theano: new features and speed improvements. NIPS 2012 deep learning workshop.
34. Theano Development Team (2016) Theano: A Python framework for fast computation of mathematical expressions. arXiv e-prints: http://arxiv.org/abs/1605.02688
35. Nickolls J., Buck I., Garland M., Skadron K. (2008) Scalable Parallel Programming with CUDA. ACM Queue, 6, 2, 40-53.
36. Gama-Castro S, Salgado H, Santos-Zavaleta A, Ledezma-Tejeida D. et al. (2016) RegulonDB version 9.0: high-level integration of gene regulation, coexpression, motif clustering and beyond. Nucleic Acids Res. 44(D1), D133-143.
37. Helmann J. (1995) Compilation and analysis of Bacillus subtilis sigma A-dependent promoter sequences: evidence for extended contact between RNA polymerase and upstream promoter DNA. Nucleic Acids Res. 23(13), 2351–2360.
38. Dreos, R., Ambrosini, G., Périer, R., Bucher, P. (2013) EPD and EPDnew, high-quality promoter resources in the next-generation sequencing era. Nucleic Acids Research, 41(Database issue): D157-64.
39. Pedregosa et al. (2011) JMLR 12, pp. 2825-2830.